\documentclass[pra,showpacs,twocolumn,superscriptaddress,floatfix]{revtex4-1}
\usepackage{epsfig}
\usepackage{amsmath}
\usepackage{amssymb}
\usepackage{amsfonts}
\usepackage{mathptmx}
\usepackage{dcolumn}
\usepackage{eucal}
\usepackage{bm}
\usepackage{color}
\usepackage[colorlinks,linkcolor=blue,citecolor=blue]{hyperref}
\usepackage{epstopdf}
\usepackage{bbold}
\usepackage{url}

\usepackage{dsfont}

\newcommand{\Time}{\mathcal{T}}
\newcommand{\Temperature}{T}

\newcommand{\average}[1]{\langle #1 \rangle}
\newcommand{\Trace}{{\rm Tr}}

\newcommand{\mathU}{\mathcal{U}}

\newcommand{\hatH}{\hat{H}}
\newcommand{\hatA}{\hat{A}}

\newcommand{\hatf}{\hat{f}}

\newcommand{\hatlambda}{\hat{\lambda}}

\newcommand{\Path}{ \mathcal{P}}

\newcommand{\Amplfinal}{ A_{\Path}^{\lambda f}}

\newcommand{\twobytwo}[4]{\left( \begin{array}{cc} #1 & #2 \\ #3 & #4 \end{array}\right)}

\def \ket#1{\mathinner{|{#1}\rangle}}
\def \bra#1{\mathinner{\langle{#1}|}}

\def\braket#1{\mathinner{\langle{#1}\rangle}}

\newcommand{\ketbra}[2]{{\mathinner{| {#1} \rangle \langle {#2} |}} }

\newcommand{\matrixel}[3]{{\mathinner{\langle{#1}| {#2} | {#3}\rangle}} }

\newcommand{\Forder}{\overrightarrow{T}}

\newcommand{\MGF}{G_\lambda}

 \newcommand{\work}{W}

\def\tmp{TMP}
\def\paper{Article}

\begin{document}

\title{Full distribution of work done on a quantum system for arbitrary initial states}

\author{P. Solinas}
\affiliation{SPIN-CNR, Via Dodecaneso 33, 16146 Genova, Italy}
\author{S. Gasparinetti}
\affiliation{Department of Physics, ETH Z\"urich, CH-8093 Z\"urich, Switzerland}

\date{\today}

\begin{abstract}
We propose a novel approach to define and measure the statistics of work, internal energy and dissipated heat in a driven quantum system.
In our framework the presence of a physical detector arises naturally and work and its statistics can be investigated in the most general case.
In particular, we show that the quantum coherence of the initial state can lead to measurable effects on the moments of the work done on the system. At the same time, we recover the known results if the initial state is a statistical mixture of energy eigenstates.
Our method can also be applied to measure the dissipated heat in an open quantum system.
By sequentially coupling the system to a detector, we can track the energy dissipated in the environment while accessing only the system degrees of freedom.
\end{abstract}

\maketitle

\section{Introduction}
One of the cornerstones of the Copenhagen interpretation of quantum mechanics is the measurement postulate: after a projective measurement, the wave function collapses into an eigenstate of the measured observable. 
In this framework, two subsequent measurements of an observable are not independent, as the first measurement perturbs the state of the system and thereby affects the result of the second \cite{dirac1967principles}.
Still, there are quantities in classical physics which are not "local" in time and need two (or more) observations to be determined. Among them are the charge flowing through and the work done on a system. In such cases, the extension of classical definitions and protocols to the quantum realm is not straightforward.
Recently, the statistics of the work done on a quantum system and, more generally, its energy exchanges have attracted much attention \cite{esposito2009nonequilibrium, campisi2011colloquium,dorner2013extracting, mazzola2013measuring, campisi2013employing, Batalho2014, roncaglia2014work, an2014experimental}. Besides a fundamental interest, the thermodynamics of quantum systems has important implications to the energetic performance of quantum devices \cite{nielsen-chuang_book} and quantum heat engines \cite{Kosloff-Levy2014}.

An established protocol to measure work involves a double \emph{projective} measurement of the energy of the system at the beginning and at the end of the evolution. Such a two-measurement protocol (TMP) can be described in terms of classical conditional probabilities \cite{kurchan2000quantum,tasaki2000jarzynski, esposito2009nonequilibrium, campisi2011colloquium}. It has proven successful in formulating quantum fluctuation relations in a setting where the system is initially in a statistical mixture of energy eigenstates.
However, \tmp{} has limitations that have so far failed to receive the due attention. These limitations become apparent when one tries to apply \tmp{} to a more general class of processes, namely, those in which the system is initially in a quantum-coherent superposition of different energy eigenstates. Most quantum gates developed in the context of quantum information and computation \cite{nielsen-chuang_book} 
belong to this class. The problem with \tmp{} is that the initial measurement forces the system into an eigenstate of the initial Hamiltonian.
The ultimate result of this operation is to reduce the dynamics to a classical statistical one \cite{Batalho2014} and to destroy the interference effects that generate, in Feynman's words, the "interfering alternatives" in the dynamics \cite{feynman1965quantum}.
In this respect, \tmp{} fails to answer in a general way the most straightforward and important question \cite{solinas2013work, salmilehto2014quantum}: how much energy is it needed in order to perform a given quantum operation on an arbitrarily prepared quantum system?

In this \paper, we address the key question we have just posed by proposing a measurement protocol that is meant to preserve the quantum-mechanical nature of the work performed on a quantum system. In this protocol, a quantum detector is coupled to the system at the beginning and at the end of the evolution. The information on the energy is stored in a phase shift that can be measured, for example, by interferometric means. Our protocol can be formally derived from a path-integral description of the dynamics by adding a constraint on the admissible paths \cite{sokolovski2005path, sokolovski2006simultaneous, sokolovski2013path}. Its predictions coincide with that of \tmp{} for a mixed initial state. However, as soon as we introduce quantum coherence in the initial state, we find a stark disagreement between the two protocols, even at the level of the first moment of the distribution, that is, the average work. We discuss the reasons for this disagreement and set the stage for further investigation. We also discuss how to extend our protocol to measure the dissipated work in driven, open quantum systems, by accessing the degrees of freedom of the system only. As compared to previous proposals relying on measurements of the environment \cite{gasparinetti2014fast,pekola2013calorimetric, campisi2013employing}, our protocol may provide an experimentally more accessible way to measure the statistics of heat and work in this case.

\section{General formalism}\label{sec:formalism}

It is known that the work done on a quantum system cannot be associated to a hermitian operator and, therefore, it is not an observable \cite{campisi2011colloquium, talkner2007fluctuation}. In general, the work performed on an open system depends on the full evolution of the system and not only on its initial and final state. As quantum trajectories (or paths) play a key role in determining energy exchanges and dissipation, we find it natural to tackle the work measurement problem by using a path integral approach \cite{feynman1965quantum}.  The formalism we describe in this section is an adaptation of that developed by Sokolovski in a series of papers \cite{sokolovski2005path, sokolovski2006simultaneous, sokolovski2013path}. We refer to the Appendix \ref{app:path_integral} and to the original papers for more technical details.

We consider a closed quantum system whose dynamics is generated by a time-dependent Hamiltonian $\hatH_S(t)$. The drive starts at $t=0$ and ends at $t=\Time$.
The corresponding unitary evolution operator can be approximated as $U(\Time) = \Forder \exp{(-i \int_0^\Time dt \hatH_S(t))} \approx \Pi_{k=0}^N  e^{-i \Delta t  \hatH_S^k} $ where $\Forder$ denotes the time-ordering product and in the second writing we have discretized the time in $N+1$ steps of length $\Delta t$ and used the notation $\hatH_S(k \Delta t) = \hatH_S^{k}$. 

\begin{figure}[t!]
\includegraphics[scale=.4]{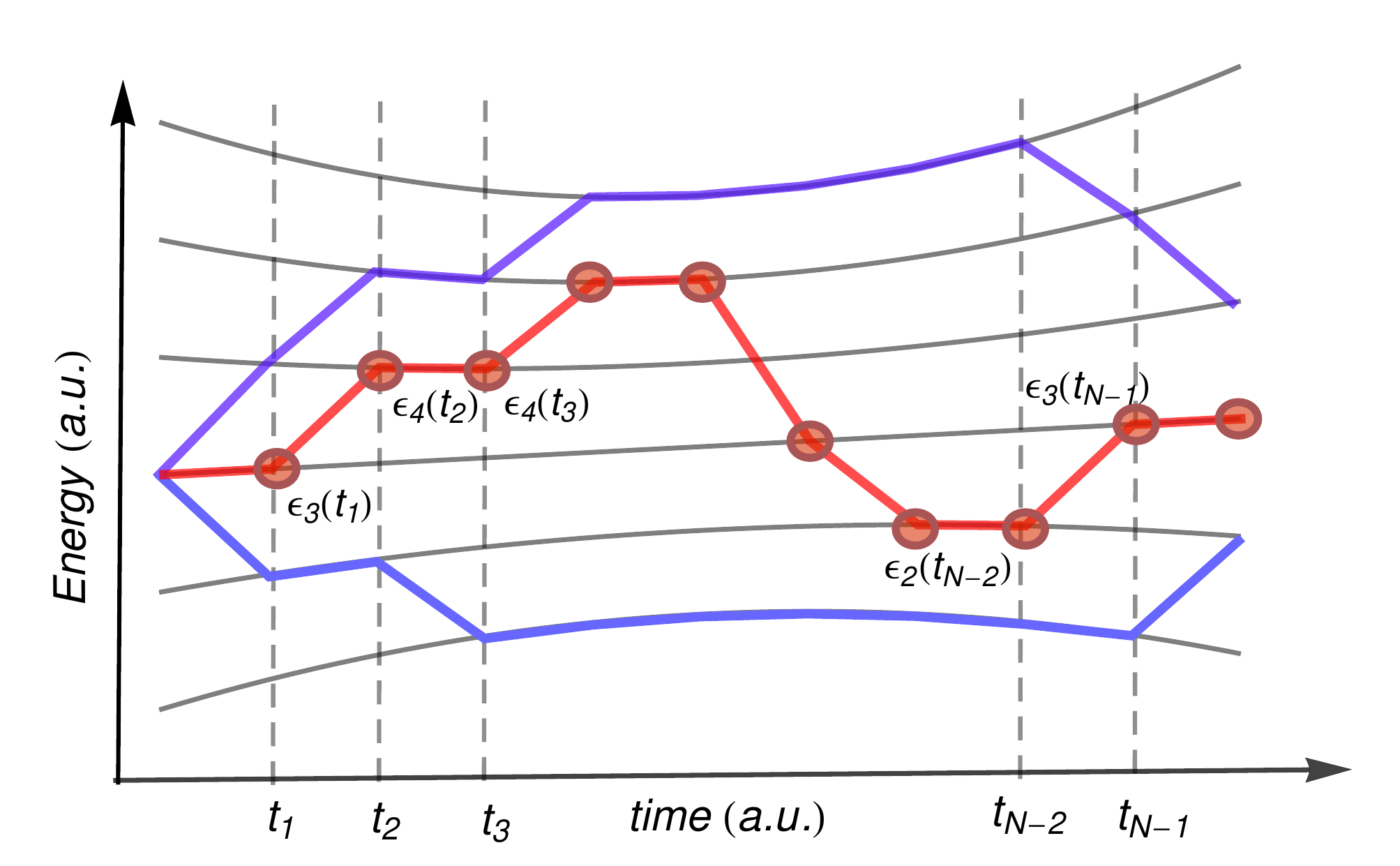}
\caption{\textbf{Quantum work and path integral.} Pictorial representation of the unitary evolution of a quantum system from the initial state $\psi_0$ (in this case an eigenstate fo the initial Hamiltonian) to the generic final state $\psi_\Time $, described in terms of paths in energy space. The time-dependent energy spectrum $\epsilon_i(t)$ of the system Hamiltonian is plotted in black. Quantum trajectories (blue, red) consists of a sequence of jumps between different eigenstates. 
The red trajectory satisfies the constraint (in this case $\Delta U=0$) while the blue ones do not.
}
\label{path_figure}
\end{figure}

The probability amplitude to go from the initial state $\ket{\psi_0}$ to a given final state $\ket{\psi_\Time}$ can be decomposed into Feynman paths \cite{feynman1965quantum}.
Differently from the usual approach, in which the dynamics is described in the position-momentum basis, we exploit the freedom to choose any complete basis at each time to decompose the paths.
If we are interested in the behavior of a time dependent operator $\hatA(t)$, the preferred basis is the one composed of its eigenstates, i.e., $\hatA(t) \ket{a_i(t)} = a_i(t) \ket{a_i(t)}$.
The idea behind this choice is that it allows us to associate to $\hatA(t)$ a value $a(t)$ depending on the path traversed during the evolution.
In a more formal way, we write $t_k = k \Delta t$ and $\{ \ket{a_i(t_k)} \} \equiv \{ \ket{a_{i_k}^k} \}$ for a complete basis set.
By inserting the completeness relation for $\ket{a_{i_k}^k}$ into the expression for the probability amplitude to go from $\ket{\psi_0}$ to $\ket{\psi_\Time}$, we obtain (see \ref{app:path_integral})
\begin{equation}
 \matrixel{\psi_\Time}{U(\Time)}{\psi_0}  \approx \matrixel{\psi_\Time }{\Pi_{k=0}^N  e^{-i \Delta t  \hatH_S^k}}{\psi_0} = \sum_{all~\Path} \mathcal{A}_\Path
\end{equation}
where $\Path$ is the path defined by the sequence of states $\{ \ket{a_{i_0}^0}, \ket{a_{i_1}^1}, ... , \ket{a_{i_N}^N} \}$ (see Fig.~\ref{path_figure}) and $\mathcal{A}_\Path$ is the probability amplitude to go from $\ket{\psi_0}$ to $\ket{\psi_\Time}$, associated to that path.
Along the path $\Path$, the operator $\hatA(t)$ takes the set of values $\left(a_{i_1}^1, a_{i_2}^2, ... , a_{i_N}^N\right) \equiv a(t)$.
Thus, we can also associate to $\Path$ any functional $F[\Path]$ of $a(t)$.

At this point, we add a constraint to the evolution by requiring $F[\Path]$ to take the value $f$.
The constrained probability amplitude reads $\mathcal{A}[f] = \sum_{\Path} \delta(F[\Path]- f) \mathcal{A}_{\Path}$.
As in \cite{sokolovski2013path}, we consider functionals of the form
$F[\Path] =  \int_0^\Time dt \beta(t) a(t) = \Delta t \sum_{k=0}^N \beta_k a_{i_k}^k$, 
where $\beta(t)$ is an arbitrary function.
The Dirac delta in the expression for $\mathcal{A}[f]$ can be written as a Fourier transform in a conjugate space described by the variable $\lambda$, as follows: $\delta(F[\Path]- f) = \int d\lambda \exp{[i \lambda( F[\Path]- f )]}$. Notice \cite{sokolovski2005path} that $\lambda$ and $f$ can be thought of as eigenvalues of conjugate operators $\hatlambda$ and $\hatf$ acting on an additional Hilbert space, their corresponding eigenstates $\ket{\lambda }$ and $\ket{ f}$ satisfying the relation $\braket{\lambda| f}= e^{-i \lambda f }$. 
Denoting with $\hatA_k \equiv \hatA(k \Delta t)$ and recalling that $\hatA_N \ket{a_{i_N}^N} = a_{i_N}^N \ket{a_{i_N}^N}$,
we can write (see Appendix \ref{app:path_integral}) $\mathcal{A}[f] = \int d \lambda  \sum_{\Path} \Amplfinal$,
where
\begin{equation}
\begin{split}
  \Amplfinal =
   &\matrixel{\psi_\Time , \lambda }{e^{-i \Delta t  (\hatH_S^N - \hatlambda \beta_N \hatA_N)} }{a_{i_N}^N}\cdot \ldots\\ 
   \cdot &\matrixel{a_{i_1}^1}{ e^{-i \Delta t  (\hatH_S^{0} - \hatlambda \beta_{0} \hatA_{0} )} }{a_{i_0}^0}
 \cdot \braket{a_{i_0}^0|\psi_0, f}\ .
\end{split}
\end{equation}
is the probability amplitude to go from the state $\ket{\psi_0,f }$ to the state $\ket{\psi_\Time , \lambda }$ \cite{sokolovski2005path, sokolovski2006simultaneous, sokolovski2013path}.
The evolution described by $\Amplfinal$ is generated by the effective Hamiltonian
\begin{equation}
 \hatH(t) =  \hatH_S(t) -\hatlambda ~ \beta(t) \hatA(t) \ .
 \label{eq:H_von_neumann}
\end{equation}
Equation \eqref{eq:H_von_neumann} plays a central role in our work and it is worth a few comments.
(i) The additional Hilbert space we introduced can be related to a detector in the von Neumann measurement scheme \cite{VonNeumann1955}. Therefore, requiring that the functional $F$ assumes the value $f$ along the evolution is equivalent to introducing a detector and coupling it to the observable we wish to measure.  Here, $\hatlambda$ and $\hat{f}$ act as the momentum and position operator of the detector, respectively.
(ii) The interaction described by \eqref{eq:H_von_neumann} does not induce any transition between the eigenstates $\ket{\lambda }$ of the detector momentum.
(iii) The information about the system-detector interaction -- and hence about the value taken by the functional $F$ -- is encoded in the phase accumulated between the eigenstates $\ket{\lambda }$ and $\ket{\lambda '}$.

Observation (iii) suggests that the statistics of the integrated observable $\hatA(t)$ can be determined by measuring the phase of the detector, as done in the full-counting-statistics approach (FCS) \cite{levitov1993charge,levitov1996electron, nazarov2003full}.
Let the composite system be initially described by the factorized density operator $\rho_0 = \rho_S^0 \otimes \rho_D^0$, where $\rho_S^0$ and $\rho_D^0$ are the density operators of the system and the detector, respectively. Then the phase difference acquired between the eigenstates $\ket{\lambda/2 }$ and $\ket{- \lambda/2 }$ of the detector reads
\begin{equation}
  \MGF = \frac{\matrixel{\lambda/2 }{\rho_D(t) }{-\lambda/2 }}{\matrixel{\lambda/2 }{\rho^0_D }{-\lambda/2 } } = \Trace_S \Big[ U_{\lambda/2}(t) \rho^0_S U_{-\lambda/2}^\dagger(t) \Big]
  \label{eq:CGF}
\end{equation}
where $U_{\lambda }(t)= \Forder \exp{[-i \int_0^t dt' (\hatH_S -\lambda \beta \hatA)]}$ is the evolution operator generated by (\ref{eq:H_von_neumann}). 
The function $\MGF$ plays the role of a moment generating function, as the $n$-th moment of $A$ is given by $\average{A^n} = (-i)^n d^n \MGF/d \lambda^n |_{\lambda=0}$ \cite{clerk2011full, bednorz2012nonclassical}.

\section{Internal energy of a closed system}

We now have the instruments to determine the variation of the internal energy of a driven closed system.
Starting from Eq. (\ref{eq:H_von_neumann}), we take $\hatA(t) = \hatH_S(t)$ and $\beta(t) = \delta(t-\Time) - \delta(t)$ \cite{sokolovski2013path}. This corresponds to coupling the detector and the system only at the beginning and at the end of the drive. 
(More precisely, we couple the system and the detector at time $t=0^-$ and at time $t=\Time^+$, i.e., immediately before and after the starting and ending drive.)
The resulting $\MGF$ is given by Eq.~(\ref{eq:CGF}), with the evolution operator (see Appendix \ref{app:path_integral})
\begin{equation}
  U_{\lambda/2}(\Time)  
  = e^{i \frac{\lambda}{2} H_{S}(\Time) } U(\Time)   e^{-i \frac{\lambda}{2} H_{S}(0)} \ .
  \label{eq:U_t}
\end{equation}
The so-obtained $\MGF$ is a measurable quantity and can be used to determine all moments of the internal-energy variation $\Delta U$ in the system. However, the interpretation of this result presents some subtleties, which we are now going to discuss.

It is known from previous work \cite{nazarov2003full, clerk2011full, bednorz2010quasiprobabilistic, bednorz2012nonclassical}
that, in general, the Fourier transform of the $\MGF$ in Eq. (\ref{eq:CGF}) cannot be associated to a probability distribution. A similar problem is encountered when defining the FCS of electron transfer across a superconducting device \cite{Belzig2001}. If a probability distribution cannot be defined for the variation of the internal energy, the question then arises what is the meaning of the moments generated by $\MGF$.
To clarify this point, let us first analyze the first moment, which for a closed system corresponds to the average work performed on the system. A physical expectation for the result can be developed by considering the following {\it gedankenexperiment}. We repeatedly prepare the system in the same initial state $\rho_S(0)$. Half of times we just measure $\hatH_S(0)$ and determine its average $\average{\hatH_S(0)}$. The remaining times we first apply the desired evolution to arrive at $\rho_S(t) = U(t)\rho_S(0) U^\dagger (t)$ and then measure $\hatH_S(\Time)$ to determine $\average{\hatH_S(\Time)}$. According to this procedure, we estimate variation of the internal energy as $\Delta U= \average{\hatH_S(\Time)} - \average{\hatH_S(0)}$. This result is the same as obtained from Eq.~(\ref{eq:CGF}); by contrast, it cannot be reproduced by \tmp{}.
To pinpoint the differences between the two methods, let us explicitly write $\Delta U$ as obtained from Eq.~(\ref{eq:CGF}) :
 \begin{equation}
 \Delta U 
 = \sum_{i_0}  \rho_{S,i_0, i_0}^0 \sum_k W_{k,i} (\epsilon_{k}^\Time - \epsilon_{i}^0)+\sum_{k, i_0 \neq j_0 }  \rho_{S,i_0, j_0}^0 \epsilon_{k}^\Time U_{k,i} U^\dagger_{j,k} \ ,
 \label{eq:W_from_G}
 \end{equation}
where $\rho_{S,i_0, j_0}^0 = \matrixel{\epsilon_{i_0}}{\rho_S^0}{\epsilon_{j_0}}$, $U_{k,i}= \matrixel{\epsilon_{k}^\Time}{ U(\Time) }{\epsilon_{i}^0}$, $ U^\dagger_{j,k} = \matrixel{\epsilon_{j}^0}{U^\dagger(\Time)}{\epsilon_{k}^\Time}$, $W_{k,i} = |\matrixel{\epsilon_{k}^\Time}{ U(\Time) }{\epsilon_{i}^0}|^2$, and $\ket{\epsilon_{k}^0}$ and $\ket{\epsilon_{k}^\Time}$ are the eigenstates of the Hamiltonian at the beginning and at the end of the evolution, respectively.
The first term in \eqref{eq:W_from_G} is the same as in \tmp{} \cite{engel2007jarzynski, esposito2009nonequilibrium, campisi2011colloquium} and can be straightforwardly interpreted in terms of classical conditional probabilities.
On the contrary, the remaining terms, which depend on the initial coherences $\rho_{S,i_0, j_0}^0$, are of a purely quantum nature. These terms are destroyed by the initial measurement of $\hatH_S(0)$ performed in TMP. 
The fact that the interfering terms can have important effect in the statistics of the work was first pointed out in Ref. \cite{solinas2013work,  Allahverdyan2014}.

The situation is well exemplified by the cyclic evolution of a coherent superposition of energy eigenstates into itself. As both the initial and final state and the initial and final Hamiltonians are the same, we would naturally expect $\Delta U=0$. However, this needs not be the case in \tmp{}, as the individual energy eigenstates after the first measurement can evolve into different states with different energies. 
As specific example we consider the case of two level system driven by a periodic Hamiltonian so that $\hatH_S(\Time)= \hatH_S(0)$.
We initialize the system in a state $\ket{\psi_0}$ that, a part from a phase factor, in left unchanged by the evolution generated by $\hatH_S(\Time)$, i.e., $ U(\Time) \ket{\psi_0}= e^{i \xi} \ket{\psi_0}$.
The existence of such a state is guaranteed, for instance, by Floquet theorem \cite{grifoni1998driven}. 
Clearly, the internal energy of the system does not change and, therefore, $ \Delta U =0$. This is correctly predicted by the first moment calculated from $\MGF$.

But, in general, $\ket{\Psi_0}$ needs not be an eigenstate of $H_S(0)$. We consider the case in which $\ket{\Psi_0}=\cos\alpha \ket{\epsilon_1}+\sin\alpha \ket{\epsilon_2}$ where $\ket{\epsilon_i}$ ($i=1,2$) are eigenstate of the initial (and final) Hamiltonian.
If we take $\alpha$ to be a free parameter, then the requirement of cyclic evolution for $\ket{\Psi_0}$ forces the evolution operator to take the form in the $\{ \ket{\epsilon_1}, \ket{\epsilon_2}\} $
$$
U(\mathcal T) = \twobytwo{\cos \xi + i \cos 2\alpha \sin \xi}
{i \sin 2\alpha \sin \xi}
{i \sin 2\alpha \sin \xi}
{\cos \xi - i \cos 2\alpha \sin \xi}.
$$
With the TMP,  after the first measurement, the system is found in $\ket{\epsilon_1}$ with probability $\cos^2\alpha$ and in $\ket{\epsilon_2}$ with probability $\sin^2\alpha$. These two states now evolve independently as the ``interfering alternatives'' have been destroyed by the projective measurement. The final result for the work distribution can be computed in terms of classical conditional probabilities $P_{ij}$ for the system to make a transition between states $i$ and $j$. In particular, for the average change in the internal energy, one finds
$$
\Delta U = \Delta E (P_{12}-P_{21}) = \Delta E \cos 2\alpha \sin^2 2\alpha \sin^2 2\xi \ ,
$$
where $\Delta E = \matrixel{\epsilon_2}{H_S(0)}{\epsilon_2}-\matrixel{\epsilon_1}{H_S(0)}{\epsilon_1}$. We have thus found that $\Delta U$ is generally nonzero, except in the cases $\xi=0$ (trivial evolution), $\alpha=0,\pi/2$ (the initial state is an energy eigenstate), and $\alpha=\pi/4$ (equal superposition of the two eigenstates).

The interpretation of higher-order moments is not trivial and stands as an open question in the field \cite{bednorz2012nonclassical}. As already mentioned, it is known from the FCS \cite{nazarov2003full, clerk2011full, bednorz2010quasiprobabilistic, bednorz2012nonclassical} that the Fourier transform of $\MGF$ is a quasi-probability which can assume negative values. A probability distribution can be retrieved in some cases after partial integration of the relevant Wigner function \cite{nazarov2003full, clerk2011full}. Ultimately, these complications are rooted in the full quantum treatment of the detector \cite{clerk2011full}. Indeed, different types of measurements performed at the end of the evolution yield different distributions for the same quantity, each of which must be interpreted accordingly. 
The measurement of the phase of the detector has the advantage that, since $\hatlambda$ is a constant of motion, preserves the ``quantumness'' of the evolution and leads to Eq.~\eqref{eq:CGF}. 
It is our belief that the quantum correlations stemming from Eq.~\eqref{eq:CGF} should not be ignored; instead, they deserve further exploration. For instance, the negativity in the quasi-probability distribution of work can be thought of as due to nonclassical temporal correlations of the energy, leading to the violation of a Leggett-Garg-type inequality \cite{Leggett1985,clerk2011full, bednorz2012nonclassical}. Further progress in this direction will hopefully appear in future work.

\section{Open system and heat statistics}

We now turn our attention to the more general case in which the system is coupled to an environment during the drive. In order to determine the work performed on the system, we need to complement the measurement of the internal energy discussed above with one of the dissipated heat. To this end, different approaches have been proposed, including the measurement of an engineered environment \cite{campisi2011colloquium, pekola2013calorimetric, gasparinetti2014fast, gasparinetti2014heat,carrega2014functional}.
Yet measuring the environment is a challenging task, restricting the applicability of these proposal to specific physical realizations. In the following, we describe an extension of our measurement protocol that allows one to obtain the statistics of the work and dissipated heat by accessing only the system degrees of freedom.

We describe the open system by the Hamiltonian $\hatH=\hatH_S + \hatH_{SE} + \hatH_E$ where $\hatH_E$ and $\hatH_{SE}$ are the environment and system-environment coupling Hamiltonians, respectively, and assume weak coupling between the system and environment.
We first take both $\hatH$ and $\hatH_S$ to be time independent and consider a measurement of $\hatH_S$. Then (\ref{eq:U_t}) simplifies into 
$U_{-\lambda/2} =  e^{- i  \frac{\lambda}{2} \hatH_S} e^{ -i  \Time \hatH} e^{ i  \frac{\lambda}{2} \hatH_S}$.
As $\hatH$ and $\hatH_S$ are constant, no external work is done on the system and the variation of internal energy must correspond to the dissipated heat. 
We can also show (see Appendix \ref{app:syst_vs_env_meas}) that the statistics obtained from the above equation is the same as the one obtained by measuring the environment degrees of freedom 
\cite{campisi2009fluctuation,esposito2009nonequilibrium,gasparinetti2014heat, silaev2014lindblad}.
We conclude that for an open system with constant Hamiltonian, the scheme gives the statistics of the dissipated heat $Q$ (Appendix \ref{app:syst_vs_env_meas}).

For a time-dependent $\hatH_S(t)$, we discretize the evolution in $N$ time intervals $\Delta t$, denote $\hatH^k = \hatH_S^k + \hatH_{SE} + \hatH_E$ with $U_{k} = e^{-i \Delta t \hatH^{k} }$. 
Within each time interval $\Delta t$, the Hamiltonian is constant. At the beginning and at the end of each interval, we instantaneously couple our detector to $\hatH_S^k$.
In analogy with \eqref{eq:U_t}, the evolution operator for each interval reads 
\begin{equation}
  \mathU _{\lambda/2} ^{k} = e^{- i  \frac{\lambda}{2} \hatH_S^k} e^{-i  \Delta t \hatH^k} e^{ i  \frac{\lambda}{2} \hatH_S^k} \ .
  \label{eq:meas_block}
\end{equation}
Each $\mathU _\lambda ^{k}$ is defined so that we keep track of the heat $Q_k$ dissipated in the time interval $(k-1) \Delta t \leq t \leq k \Delta t $. As a result, the information on the dissipated heat along the evolution is stored in the phase of the detector. 
Notice the opposite sign in the exponents with respect to Eq. (\ref{eq:H_von_neumann}) takes into account the fact that an emission (absorption) by the environment, i.e., decreasing (increasing) of the environment energy, corresponds to an absorption (emission) process of the system, i.e., increasing (decreasing) of the system energy.

\begin{figure}[t!]
\includegraphics[scale=.5]{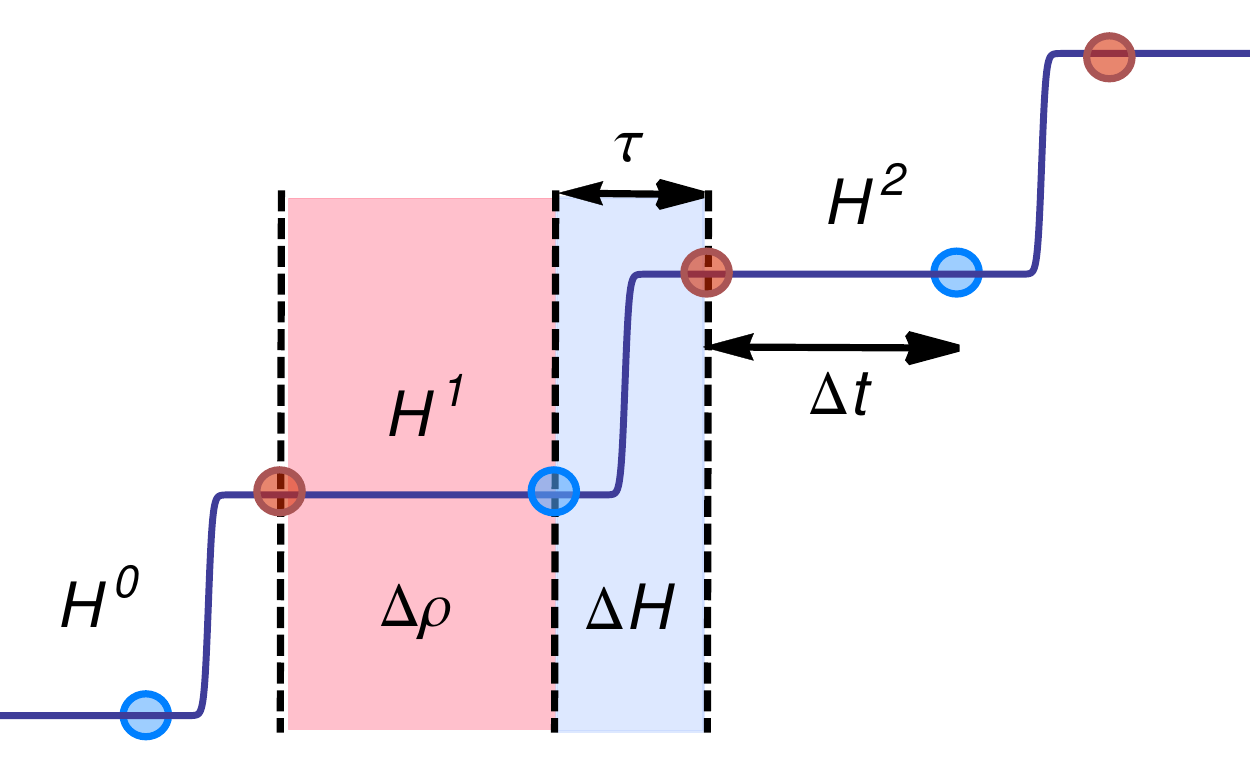}
\caption{\textbf{Measuring work and dissipation in open quantum systems.} Schematic representation of the sequence of driven evolutions and interactions with the detector with the open system protocol in Eq. (\ref{eq:U_full_work}).
The evolution steps $\exp{(- i \Delta t \hatH^k)}$ are represented by the flat line and characterized by the Hamiltonian $H^k$. Each coupling with the detector is represented by a dot. The coupling is either of the form $\exp{(-i \lambda  \hatH_{S}^k )}$ (red dots), or $\exp{(i \lambda \hatH_{S}^k )}$ (blue dots). In the blue-shadowed region, the evolution is frozen ($\Delta\rho = 0$) and the Hamiltonian changes by $\Delta H$.  In the red-shadowed region, the Hamiltonian is constant ($\Delta H = 0$) while the density operator changes by $\Delta \rho$.} 
\label{meas_figure}
\end{figure}

In order to account for the variation of the internal energy as well, we must add a measurement of $H_S$ at the beginning and end of the evolution (Appendix \ref{app:first_moment}).
Putting things together, the total evolution operator reads
\begin{equation}
 U_{\lambda/2} = e^{i  \frac{\lambda}{2} \hatH_S^N} \Pi_{k=0}^{N} \mathU_{\lambda/2} ^{k} e^{-i  \frac{\lambda}{2} \hatH_S^0}. 
 \label{eq:U_full_work}
\end{equation}
A pictorial representation of the scheme described by \eqref{eq:U_full_work} is presented in Fig.~\ref{meas_figure}.
In the case of unitary evolution, $\hatH^k = \hatH_S^k$ and we immediately recover the closed-system result for the variation of the internal energy.
The moment generating function is the same as in (\ref{eq:CGF}) with $U_{\lambda}$ given by (\ref{eq:U_full_work}).
Let us calculate its first moment, which gives the average work
$ \work=-i d \MGF/d \lambda|_{\lambda=0}$.
We find (Appendix \ref{app:first_moment})
\begin{equation}
  \work =  \Trace_{S} \Big[ \hatH_S^{N} \rho_{S,N} - \hatH_S^{0} \rho_{S,0} - \Sigma_{k} \hatH_S^{k}  \Delta \rho_{S,k} \Big] ,
    \label{eq:W_meas}
\end{equation}
where $\Delta \rho_{S,k} = \rho_{S,k}- \rho_{S,k-1}$.
In the first two terms we recognize the variation of the internal energy of the system: $\Delta U = \Trace_{S} \Big[ \hatH_S^{N} \rho_{S,N} - \hatH_S^{0} \rho_{S,0} \Big]$.
Accordingly, we identify the remaining term with the dissipated heat: 
$Q =  \Sigma_{k} Q_k =  \Sigma_{k} \Trace_{S} \Big[  \hatH_S^{k} \Delta \rho_{S,k} \Big]$.
Notice that while $\Delta U$ depends only on the initial and final state of the system, $Q$ is determined by the full dissipative evolution as in the classical counterpart (Appendix \ref{app:first_moment}).

In the fast-decoherence limit, the dissipated heat takes an illuminating form.
When energy-relaxation processes are much faster than the dynamics induced by the drive, we always find the system in its instantaneous thermal equilibrium state $\rho_S(t) = \exp{[- \hatH_S(t)/k_B \Temperature]}/Z_S(t)$ where $Z_S(t)$ is the partition function of the system and $\Temperature$ is the temperature of the environment. In other words, the system evolves through states of quasi-equilibrium.
Defining the Von Neumann entropy as $S = - \Trace [ \rho_S \log \rho_S]$, we can show (see Appendix \ref{app:entropy}) that the variation of the density operator is related to it by $\Trace_S \Big[  \hatH_S^k \Delta \rho_k \Big]  = k_B T  \Delta S_k$ where $\Delta S_k$ is the variation of entropy at time $t_k$.
Then we can link the variation of entropy to the dissipated heat by the relation $Q_k= k_B T  \Delta S_k$, confirming the above interpretation of $Q$ as the dissipated heat.

There is an alternative way to interpret Eq.~(\ref{eq:W_meas}).
Taking the time derivative of the average internal energy, we have $d \average{\hatH_S (t)} /dt = \average{\dot{\hatH}_S (t) \rho(t)} +  \average{\hatH_S (t) \dot{\rho}(t)}$ \cite{solinas2013work}. If the evolution is unitary, the second contribution vanishes and we can relate the variation of the system Hamiltonian to the instantaneous work done on the system.
By expanding the product in \eqref{eq:U_full_work}, we identify pairs of sequential system-detector interactions of the form $\exp{(i \lambda  \hatH_{S}^{k+1}/2)}\exp{(-i \lambda  \hatH_{S}^{k}/2)}$. Each such pair effectively keeps track of a variation in the Hamiltonian. As the variation is instantaneous, the system has no dynamics. We can interpret the action of the pairs as a ``measurement'' of the work done on the system by an external force.
This interpretation is strengthened by the analysis of Eq.~(\ref{eq:W_meas}). By regrouping the terms, we can write it as $ W = \Trace_{S} \Big[ \Sigma_{k} \Delta \hatH_S^{k}   \rho_{S,k} \Big]$
where $\Delta \hatH_S^{k} = \hatH_S^{k+1} -\hatH_S^{k}$.

One may wonder whether the repeated coupling to the detector can 'freeze' the dynamics of the system (dynamic Zeno effect).
This turns out not to be the case: a dynamic Zeno effect would require $\lambda \to \infty$, while we derive our physical quantities, i.e., the moments of the work done, in the opposite limit $\lambda \to 0$.
Our protocol can instead be regarded as a noninvasive measurement \cite{Gardiner2004, bednorz2012nonclassical} of the work distribution.
In fact, the moments generated by $G_\lambda$ depend on evolution operators that describe the dynamics of the open system without a detector.

\section{Conclusions and outlook}

In summary, we have shown that the statistics of work performed on a quantum system exhibits nonclassical correlations in a deeper and more fundamental way that it had so far been appreciated. Such correlations become apparent once one replaces the customary double projective measurement with a less-invasive coupling to a quantum detector. The resulting protocol is immediately applicable to the case of unitary evolution and can be suitably extended to treat open quantum systems. Our approach puts the problem of work under a new perspective and leads the way toward further investigations. In particular, the links between quantum-mechanical work, Leggett-Garg-type inequalities \cite{Leggett1985,clerk2011full}, weak measurements \cite{Gardiner2004}, and stochastic quantum trajectories \cite{Carmichael1993,Murch2013,Weber2014}, await to be fully elucidated. An experimental test of our predictions is in reach of state-of-the art quantum technology. Among different architectures, superconducting quantum circuits in combination with nearly-quantum-limited parametric amplifiers are a first choice, given the high degree on control achieved in recent experiments \cite{Vijay2012, Murch2013, Hatridge2013,Riste2013}.

\begin{acknowledgments}
We gratefully acknowledge A. Braggio and M. Carrega for fruitful discussions. 
P.S. has received funding from the European Union FP7/2007-2013 under REA
grant agreement no 630925 -- COHEAT and from MIUR-FIRB2013 -- Project Coca (Grant
No.~RBFR1379UX). S.G. acknowledges financial support from the Swiss National Science Foundation (SNF) Project 150046.
\end{acknowledgments}


\newpage

\appendix
\newpage

\section{Probability amplitude in the path integral representation}
\label{app:path_integral}

We consider a closed quantum system whose dynamics is generated by a time-dependent Hamiltonian $\hatH_S(t)$. The drive starts at $t=0$ and ends at $t=\Time$.
The corresponding unitary evolution operator is 
\begin{equation}
 U(\Time)  = \Forder e^{-i \int_0^\Time  dt \hatH_S(t)} \approx \Pi_{k=0}^N  e^{-i \Delta t  \hatH_S^k} 
\label{eq_app:U_dyn}
\end{equation}
where $\Forder$ denotes the time-ordering product and in the second writing we have discretized the time in $N+1$ steps of length $\Delta t$ and used the notation $\hatH_S(k \Delta t) = \hatH_S^{k}$. 
Our goal is to write the probability amplitude to go from $\ket{\psi_0}$ to $\ket{\psi_\Time }$ in terms of Feynman paths  \cite{feynman1965quantum}.
Equation (\ref{eq_app:U_dyn}) is approximated to the order $\Delta t ^2$ and it is a convenient way to describe the evolution in terms of path integral.
As a final step, we will take the limit $\Delta t \rightarrow 0$ to recover the usual continuous description of the evolution.

{\it At any time} $t_k$ we can find a basis $\{ \ket{a_i(t_k)} \} \equiv \{ \ket{a_{i_k}^k} \}$ such that the completeness relation $ \sum_{i_k} \ketbra{a_{i_k}^k}{a_{i_k}^k} = \mathds{1}$ holds.
In this notation, $k$ is a {\it time} index and $i_k$ denotes the eigenstate basis index at time $t_k$.
By inserting the completeness relation for $\ket{a_{i_k}^k}$ into the expression for the probability amplitude we obtain
\begin{eqnarray}
 \matrixel{\psi_\Time }{U(\Time) }{\psi_0} &=& \sum_{k=0}^N \sum_{i_k} 
 \matrixel{\psi_\Time }{e^{-i \Delta t  \hatH_S^N}}{a_{i_N}^N} \matrixel{a_{i_N}^N}{e^{-i \Delta t  \hatH_S^{N-1}}}{a_{i_{N-1}}^{N-1}} ... \nonumber \\
 && \matrixel{a_{i_2}^2}{e^{-i \Delta t  \hatH_S^{1}}}{a_{i_1}^1} \matrixel{a_{i_1}^1}{e^{-i \Delta t  \hatH_S^{0}}}{a_{i_0}^0}
 \braket{a_{i_0}^0|\psi_0}.
 \label{eq:path_int_prob_ampl}
\end{eqnarray}
The term $\matrixel{a_{i_k}^k}{e^{-i \Delta t  \hatH_S^N}}{a_{i_{k-1}}^{k-1}}$ is the probability amplitude to go from $\ket{a_{i_{k-1}}^{k-1}}$ to $\ket{a_{i_k}^k}$.
Then,
\begin{eqnarray}
\mathcal{A}_\Path &=&  \matrixel{\psi_\Time }{e^{-i \Delta t  \hatH_S^N}}{a_{i_N}^N} \matrixel{a_{i_N}^N}{e^{-i \Delta t  \hatH_S^{N-1}}}{a_{i_{N-1}}^{N-1}} ... \nonumber \\ 
  && 
 \matrixel{a_i^2}{e^{-i \Delta t  \hatH_S^{1}}}{a_{i_1}^1} \matrixel{a_{i_1}^1}{e^{-i \Delta t  \hatH_S^{0}}}{a_{i_0}^0}
 \braket{a_{i_0}^0|\psi_0} 
 \label{eq:path_amplitude}
\end{eqnarray}
is the probability amplitude to go from $\ket{\psi_0}$ to $\ket{\psi_\Time }$ passing through the sequence of states:
$\ket{a_{i_0}^0}, \ket{a_{i_1}^1}, ... , \ket{a_{i_N}^N}$.
This sequence define a {\it path} $\Path$ in the basis space $\{ \ket{a_{i_k}^k} \}$. We interpret (\ref{eq:path_int_prob_ampl}) as the sum over all the possible {\it paths} of the probability amplitudes:
\begin{equation}
 \matrixel{\psi_\Time }{U(\Time) }{\psi_0}  = \sum_{all~\Path} \mathcal{A}_\Path.
 \label{eq:A_path}
\end{equation}
In the limit $\Delta t \rightarrow 0$, we obtain a continuous path $a(t)$.
In this way, we can associate to the path $\Path$ a physical quantity $F[\Path]$ depending on it.
Mathematically, $F$ is then a functional of $\Path$.

We now add a constraint and select only the paths that satisfy some properties. 
We are asking which is the probability amplitude $\mathcal{A}[f]$ to go from $\ket{\psi_0}$ to $\ket{\psi_\Time }$ though a path $\Path$ determined by $a(t)$ for which the functional $F[\Path]$ assumes the values $f$.
The constrained probability amplitude reads
\begin{eqnarray}
  \mathcal{A}[f] = \sum_{\Path} \delta(F[\Path]- f) \mathcal{A}_{\Path}
\end{eqnarray} 
where the delta function restricts the admissible paths to those for which $F[\Path] = f$.
$F[\Path]$ could a generic functional of $\Path$.
However, as in Ref. \cite{  sokolovski2013path}, we assume that the functional depends on the integral of the path $a(t)$ (or $\ket{a_{i_1}^1}, \ket{a_{i_2}^2}, ... , \ket{a_{i_N}^N}$ in the discretized expression)
\begin{equation}
  F[\Path] =  \int_0^\Time  dt \beta(t) a(t) =  \Delta t \sum_{k=0}^N \beta_k a_{i_k}^k
\end{equation}
where $\beta(t)$ is an arbitrary function.

By writing the Dirac delta in terms of Fourier Transform  $\delta(F[\Path]- f) = \int d \lambda~\exp{[- i \lambda  ( f- F[\Path])]}$
inserting it in the path integral representation, and splitting the term $F[\Path]$ with respect to the corresponding time interval, we obtain
\begin{eqnarray}
  \mathcal{A}[f] &=& \int d \lambda~e^{-i \lambda  f } \sum_{\Path} 
  \matrixel{\psi_\Time }{e^{-i \Delta t  \hatH_S^N} e^{i \Delta t  \lambda  \beta_N a_{i_N}^N}}{a_{i_N}^N} ... \nonumber \\
  &&\matrixel{a_{i_1}^1}{ e^{-i \Delta t  \hatH_S^{0}} e^{i \Delta t  \lambda  \beta_{0} a_{i_0}^0} }{a_{i_0}^0}
 \braket{a_{i_0}^0|\psi_0}.
\end{eqnarray}
We must choose the time-dependent basis set $\{ \ket{a_{i_k}^k} \}$ considering the observable $\hatA(t)$ we are interested in.
In particular, we must take them in order that $\hatA_k \ket{a_{i_k}^k} = a_{i_k}^k \ket{a_{i_k}^k}$.
For small $\Delta t$ \cite{kleinert2009path}, we can write
\begin{eqnarray}
  \mathcal{A}[f] 
  &\approx& \int d \lambda~e^{-i \lambda  f } \sum_{\Path} 
  \matrixel{\psi_\Time }{e^{-i \Delta t  (\hatH_S^N - \lambda  \beta_N \hatA_N)} }{a_{i_N}^N} ... \nonumber \\ 
  && \matrixel{a_{i_1}^1}{ e^{-i \Delta t  (\hatH_S^{0} - \lambda  \beta_{0} \hatA_{0} )} }{a_{i_0}^0}
 \braket{a_{i_0}^0|\psi_0} \nonumber \\
  &=& \int d \lambda ~e^{-i \lambda  f } \sum_{\Path} \mathcal{A}_\Path^{\lambda }.
  \label{app:eq_A_f} 
\end{eqnarray}
Therefore, the constrained amplitude probability $ \mathcal{A}[f]$ can be written as the sum of the path amplitudes $\mathcal{A}_\Path^{\lambda }$, which are generated by the effective Hamiltonian $\hatH_S(t) - \lambda  \beta(t) \hatA(t)$.
By defining the corresponding unitary operator as
 \begin{equation}
 U_{\lambda } (t) = \Forder e^{-i  \int_0^t dt' [\hatH_S(t) - \lambda  \beta(t) \hatA(t)]},
 \label{app:eq_U_lambda}
\end{equation}
we have 
\begin{equation}
  \sum_{\Path} e^{i \lambda  F[\Path]} \mathcal{A}_\Path=  \sum_{\Path} \mathcal{A}_\Path^{\lambda } =   \matrixel{\psi_\Time }{U_{\lambda } (t)}{\psi_0}.
  \label{eq:A_p_relation}
\end{equation}

We can go further with the interpretation of the constraint. 
The parameter $\lambda$ and $f$ can be thought of as eigenvalues of conjugate operators $\hatlambda$ and $\hatf$ satisfying the relation $\braket{\lambda| f}= e^{-i \lambda f }$. 
Using the relation $ e^{-i \Delta t  (\hatH^{k} - \lambda  \beta_{k} \hatA_{k} )}  \bra{\lambda } = \bra{\lambda } e^{-i \Delta t  (\hatH^{k} - \hatlambda \beta_{k} \hatA_{k} )}  $, we can write
\begin{eqnarray}
  \mathcal{A}[f]
  &=&   \int d \lambda \sum_{\Path} 
   \matrixel{\psi_\Time , \lambda }{e^{-i \Delta t  (\hatH^N - \hatlambda \beta_N \hatA_N)} }{a_{i_N}^N} ... \nonumber \\
&&\matrixel{a_{i_1}^1}{ e^{-i \Delta t  (\hatH^{0} - \hatlambda \beta_{0} \hatA_{0} )} }{a_{i_0}^0}
 \braket{a_{i_0}^0|\psi_0, f}
 \label{app:eq_Af}
\end{eqnarray}
Based on Eq. (\ref{app:eq_Af}), we can interpret $\mathcal{A}[f]$ is the probability amplitude to go from the state $\ket{\psi_0, f}$ to the state $\ket{\psi_\Time , \lambda }$, where $\hatlambda$ and $\hat{f}$ are conjugate operator acting on an additional Hilbert space.
The latter is interpreted as {\it the Hilbert space of the detector needed to measure the special observable} \cite{  sokolovski2005path,   sokolovski2006simultaneous,   sokolovski2013path}.
The effective Hamiltonian describing the system and a quantum detector dynamics is
\begin{equation}
  \hatH(t) =  \hatH_S(t) - \hatlambda~ \beta(t) \hatA(t).
  \label{app:eq_H_tot}
\end{equation}
The approach outlined above applies to any time-dependent observable $\hatA(t)$.
In this work we take $\hatA(t)$ to be the time-dependent Hamiltonian. The power operator considered in Ref.~\cite{solinas2013work} would be another meaningful choice.

To determine the variation of the internal energy, we take $\hatA(t) = \hatH_S(t)$ and $\beta(t) = \delta(\Time-t) -\delta(t)$. When using a discretized evolution, we assume that $\delta(t_k-t)= 1/\Delta t$ for $t_k  \leq t \leq t_{k+1}= t_k+\Delta t$ and $0$ elsewhere.
From Eqs. (\ref{app:eq_U_lambda}) we have, in the limit of $\Delta t \rightarrow 0$ \cite{sokolovski2013path, sokolovski2006simultaneous}
\begin{eqnarray}
 U_{\lambda } (\Time)  &\approx&  e^{-i \Delta t  (\hatH^N - \lambda  \beta_N \hatA_N)} e^{-i \Delta t  (\hatH^{N-1} - \lambda  \beta_{N-1} \hatA_{N-1})}... 
 e^{-i \Delta t  (\hatH^0 - \lambda  \beta_0 \hatA_0)} \nonumber \\
 &\rightarrow & e^{i    \lambda  \hatH(\Time) }  U(\Time)   e^{-i    \lambda  \hatH(0)}.
 \label{eq_app:U_lambdap}
\end{eqnarray}
Therefore, the total unitary evolution corresponds to two fast couplings with the detector with an central driven evolution of the system.
From this we immediately arrive to the moment generating function $\MGF$ discussed in the main text.


\section{System versus environment measurement.}
\label{app:syst_vs_env_meas}

In the main text we have discussed how to measure the dissipated heat statistics though the system degrees of freedom. Here we show that this statistics is the same as we would obtain by measuring directly the environment \cite{  campisi2011colloquium,   gasparinetti2014heat}.

We make the standard assumption that the system and the environment are weakly coupled.
This allows us to neglect the energy related to system-environment coupling Hamiltonian.
We consider $\hatH_S$ to be time-independent as in the fundamental interaction block discussed in the main text.
The total Hamiltonian reads $\hatH = \hatH_S + \hatH_E +\hatH_{SE}$.
If we measure the degrees of freedom of the environment, we obtain a $\bar{G}_\lambda$  that has the form of  $\MGF$ 
with  $\bar{U}_{\lambda }(\Time)  = \exp{( i \lambda  \hatH_E/2)} \exp{(- i  \hatH T)} \exp{(- i \lambda  \hatH_E/2)}$ \cite{  gasparinetti2014heat}.

In the weak coupling limit, $\exp{( i \lambda  \hatH_E/2)} = \exp{[ i \lambda  ( \hatH - \hatH_S -\hatH_{SE})/2]} \approx \exp{[ i \lambda  ( \hatH - \hatH_S)/2]}$ and $[\hatH, \hatH_S] \approx 0$.
Therefore,
\begin{eqnarray}
 \bar{U}_{\lambda } (\Time) & \approx & e^{- i \lambda  \hatH_S/2} e^{ i \lambda  \hatH/2 } e^{- i  \hatH T} e^{- i \lambda  \hatH/2 } e^{ i \lambda  \hatH_S/2} \nonumber \\
 &=&  e^{- i \lambda  \hatH_S/2}  e^{- i  \hatH T}  e^{ i \lambda  \hatH_S/2} =  U_{-\lambda } (\Time).
 \label{app:eq_U_lambda}
\end{eqnarray}
From Eq. (\ref{app:eq_U_lambda}) it follows that $\bar{G}_\lambda= G_{-\lambda}$ and the statistics generated by measuring $\hatH_E$ is equal to the one obtained by measuring $\hatH_S$ 
with opposite sign.
The opposite sign in the exponents with respect to Eq. (\ref{app:eq_U_lambda}) takes into account the fact that an emission (absorption) by the environment, i.e., decreasing (increasing) of the environment energy, corresponds to an absorption (emission) process of the system, i.e., increasing (decreasing) of the system energy.

\section{First moment of the work done on a quantum system}
\label{app:first_moment}

We first restrict our attention to the dissipated heat and calculate the first moment of the moment generating function $\MGF$
\begin{equation}
  -Q=-i \frac{ d \MGF }{d \lambda} \Big|_{\lambda=0}=  \Trace_{S+E} \Big[ \frac{ d U_{\lambda }}{d \lambda} \rho_0 U_{-\lambda }^\dagger + 
  U_{\lambda } \rho_0 \frac{ d U_{-\lambda }^\dagger}{d \lambda}
    \Big] \Big|_{\lambda=0}
    \label{eq:first_moment}
\end{equation}
where $U_{\lambda } =  \Pi_{k=0}^{N} \mathU _\lambda ^{k}$, $\mathU _\lambda ^{k} = e^{- i  \lambda  \hatH_S^k/2} e^{-i  \Delta t \hatH^k} e^{ i  \lambda  \hatH_S^k/2}$, and we follow the convention that the heat flowing into the system is given a positive sign \cite{  campisi2011colloquium}.

The first term in Eq.~\eqref{eq:first_moment} reads
\begin{eqnarray}
 -i \frac{ d U_{\lambda }}{d \lambda}  \Big|_{\lambda=0} = -\frac{1}{2}&\Big[&
 \hatH_S^{N} U - U_{N} (\hatH_S^{N} -\hatH_S^{N-1}) U_{N-1}...U_{0} + ... \nonumber \\
 &&- U_{N}...U_{1} (\hatH_S^{1} -\hatH_S^{0}) U_{0} -  U \hatH_S^0
 \Big] \ .
 \label{eq:app_dU}
\end{eqnarray}
where we used the compact notation $U_{k} = e^{-i \Delta t \hatH^{k} }$.
In an analogous way, the second term reads
\begin{eqnarray}
 -i \frac{ d U_{-\lambda }^\dagger}{d \lambda}  \Big|_{\lambda=0} &=& \frac{1}{2} \Big[
 \hatH_S^{0} U^\dagger + U_{0}^\dagger (\hatH_S^{N} -\hatH_S^{N-1}) U_{1}^\dagger...U_{N-1}^\dagger +... \nonumber \\
 &+& U_{0}^\dagger...U_{N-1}^\dagger (\hatH_S^{1} -\hatH_S^{0}) U_{N}^\dagger -  U^\dagger \hatH_S^N
 \Big]\ .
\end{eqnarray}
Putting everything together in Eq.~(\ref{eq:first_moment}) and using the cyclic property of the trace, it is possible to simplify some of the evolution operators $U_{k}$.
After defining $\rho_k = U_{k}...U_{0} \rho_0 U_{0}^\dagger...U_{k}^\dagger$ and $\tilde{\rho}_0 =U_{0} \rho_0 U_{0}^\dagger$, we have
\begin{equation}
  - Q=  \Trace_{S+E} \Big[
      - \hatH_S^{N} (\rho_{N-1}- \rho_{N-2})  ... -\hatH_S^{0} (\tilde{\rho}_{0}- \rho_{0})
  \Big].
  \label{eq:app_Q1}
\end{equation}
The trace over the system and environment can be separated by observing that $\Trace_{S+E} \Big[ \hatH_S^{N} (\rho_{N}- \rho_{N-1}) \Big] = \Trace_{S} \Big[ \hatH_S^{N} \Trace_{E} \Big(\rho_{N}- \rho_{N-1}\Big) \Big] = \Trace_{S} \Big[ \hatH_S^{N} (\rho_{S,N}- \rho_{S,N-1}) \Big]$. Therefore, the dissipated heat written in terms of the system degrees of freedom reads
\begin{equation}
  Q =   \Trace_{S} \Big[ \Sigma_{k} H_S^{k} (\rho_{S,k}- \rho_{S,k-1})  \Big] = \Sigma_{k} \Trace_{S} \Big[  H_S^{k} \Delta \rho_{S,k} \Big] =  \Sigma_{k} Q_k
    \label{eq:Q_meas}
\end{equation}
where $Q_k $ is the dissipated heat in the time interval $t_{k-1} \leq t \leq t_k$.

Let us now introduce a coupling between the system and the detector at the beginning and at the end of the evolution. The unitary operator then reads 
\begin{equation}
 U_{\lambda } = e^{i  \hatlambda \hatH_S^N/2} \Pi_{k=0}^{N} \mathU _\lambda ^{k} e^{-i  \hatlambda \hatH_S^0/2}.
  \label{app_eq:U_full_work}
\end{equation}
Accordingly, the calculation in Eq.~(\ref{eq:app_dU}) is modified as
\begin{equation}
 -i \frac{ d U_{\lambda }(t)}{d \lambda}  \Big|_{\lambda=0} 
 =\hatH_S^{N} U - U \hatH_S^0 -i \frac{ d   }{d \lambda}\Big( \Pi_{k=0}^{N} \mathU _\lambda ^{k}\Big) \Big|_{\lambda=0} \ ,
 \end{equation}
which differs from Eq.~(\ref{eq:app_dU}) by the addition of the term $\hatH_S^{N} U - U \hatH_S^0$.

We find that the average work $\work$ is 
\begin{equation}
  \work 
   =  \Trace_{S} \Big[ H_S^{N} \rho_{S,N} - H_S^{0} \rho_{S,0} - \Sigma_{k} H_S^{k} (\rho_{S,k}- \rho_{S,k-1})  \Big]
    \label{eq:app_W_meas}
\end{equation}
which differs from Eq. (\ref{eq:Q_meas}) by the variation of the internal energy $\Delta H = \Trace_{S} \Big[ H_S^{N} \rho_{S,N} - H_S^{0} \rho_{S,0}\Big]$. Therefore, we have obtained the usual result $\work = \Delta H - Q$ \cite{  campisi2011colloquium}.

The heat contributions $Q_k$ in \eqref{eq:Q_meas} are related to the variation of the system density operator $\Delta \rho_k$ during infinitesimal evolutions generated by constant Hamiltonians. We can check that if the evolution is unitary, i.e., $H^k = H_S^k$, then the $Q_k$ vanish and no heat is dissipated. In fact, we have $\rho_{S,k} = U_k \rho_{S,k-1} U_k^\dagger$, $[H_{S,k},U_k ]=0$, and 
\begin{eqnarray}
 Q_k&=& \Trace_{S} \Big[ H_S^{k} \rho_{S,k}- H_S^{k} \rho_{S,k-1}  \Big] \nonumber \\ 
 &=& \Trace_{S} \Big[ U_k H_S^{k} \rho_{S,k-1} U_k^\dagger - H_S^{k} \rho_{S,k-1}  \Big] =0 \ .
\end{eqnarray}

The interpretation of the $Q_k$ as the dissipative contribution to the dynamics is strengthened by the following observation. The dissipated heat depends on the variation of the density operator $\Delta \rho_{S,k}$, which, in turn, can be due to both unitary and non-unitary dynamics. However, the unitary contribution to the change of $\rho$, i.e., the one given by $[H_S,\rho_S]$, vanishes identically when we calculate $H_S^{k} (\rho_{S,k}- \rho_{S,k-1})$. Thus, the $Q_k$ are related solely to the dissipative dynamics.
This result is analogous to the one obtained in Ref.~\cite{  solinas2013work}.

The expressions for $Q$ and $W$ can be written in another meaningful way as follows.
Instead of grouping $\Delta \rho_k$, we can keep the terms $\Delta H_S^k = \hatH_S^k - \hatH_S^{k-1}$ as written in Eq.~(\ref{eq:app_dU}). Then the dissipated heat in Eq.~(\ref{eq:app_Q1}) reads
\begin{equation}
 Q=\Trace_{S} \Big[ -H_S^{N} \rho_{S,N} + H_S^{0} \rho_{S,0} + \Sigma_{k} \Delta H_S^{k} \rho_{S,k} \Big].
\end{equation}
As the contributions of the initial and final measurements are the same, we have that
\begin{equation}
    \work 
   = \Trace_{S} \Big[  \Sigma_{k} \Delta H_S^{k} \rho_{S,k}\Big].
\end{equation}
This confirms the interpretation discussed in the main text that the work can be seen as the instantaneous energy injected in the system due to the variation of the Hamiltonian in time.

\section{Work, heat and entropy in quantum system and dynamics}
\label{app:entropy}

The Von Neumann entropy in a quantum system is defined as $S = - \Trace [ \rho_S \log \rho_S]$.
Writing it in the basis $\{ | i \rangle\}$ in which $\rho_S$ is diagonal, we obtain $ S = - \sum_i \rho_{_S, ii} \log \rho_{_S, ii}$.
If we take the time derivative of the entropy, we have 
\begin{equation}
  \frac{dS}{dt} = - \sum_i (\dot{\rho}_{S, ii} \log \rho_{S, ii} + \dot{\rho}_{S, ii}) =   - \sum_i \dot{\rho}_{S, ii} \log \rho_{ii} 
  \label{eq:dS}
\end{equation}
where the last equation comes from the fact that $\sum_i \dot{\rho}_{S, ii} = 0$ due to trace conservation.

If the decoherence time is smaller than all the system time-scale, the system is always in the (time-dependent) thermalized state. 
Under this hypothesis the above equation should be rewritten explicitly with the time-dependence $\rho_S(t) = \exp{[-\beta \hatH_S(t)]}/Z_S(t)$ where $\beta = 1/(k_B T)$ is the inverse temperature of the environment and $Z_S(t)$ is the partition function. 
In addition, we have $\log[ \rho_S(t)] = -\hatH_S(t) - \log [Z_S(t)]$.
Again, this must be intended in terms of component in the basis in which $\rho_S$ and $\hatH_S$ are diagonal.
If $\epsilon_i$ is the energy of the state $|i \rangle$, we have 
$\rho_S = \sum_i \rho_{S, ii} |i\rangle \langle i | =  \sum_i e^{-\beta \epsilon_i}/Z_S |i\rangle \langle i | $ and 
$ \log \rho_{S, ii} = -\beta \epsilon_i - \log [Z_S(t)]$.
With Eq. (\ref{eq:dS}), we can write
\begin{equation}
  \frac{dS}{dt} =     \sum_i [ \beta \dot{\rho}_{ii} \epsilon_i +  \log Z_S(t) \dot{\rho}_{S,ii} ] =   \frac{1}{k_B T}  \sum_i   \dot{\rho}_{S, ii} \epsilon_i .
\end{equation}

Keeping in mind that $H_S= \sum_k \epsilon_k |k\rangle \langle k|$, we have that $\sum_i [  \dot{\rho}_{S, ii} \epsilon_i ]  = \Trace [\frac{d\rho_S}{dt} H_S ] $ and  we can rewrite the above equation as 
\begin{equation}
 \Trace \Big[ \frac{d\rho_S}{dt} H_S \Big]= T \frac{dS}{dt}.
   \label{eq:entropy_mapping}
\end{equation}
The variation of the entropy can be related, as in the classical case, to the dissipated heat. With this identification we find  $Q_k = \Trace_S \Big[  \hatH_S^k \Delta \rho_k \Big]  = k_B T  \Delta S_k$.





\end{document}